\documentclass[twocolumn,showpacs,preprintnumbers,a4,prb]{revtex4}
\usepackage{graphicx}
\usepackage{dcolumn}
\usepackage{amsmath}
\usepackage{amssymb}
\usepackage{bm}

\begin{document}

\title{Stray-fields based features observed for low and high magnetic fields in Ni$_{80}$Fe$_{20}$-Nb-Ni$_{80}$Fe$_{20}$ trilayers}

\author{D. Stamopoulos, \footnote[1]{Author to whom any correspondence should be
addressed: densta@ims.demokritos.gr} E. Manios, N. Papachristos and I. Aristomenopoulou}

\affiliation{Institute of Materials Science, NCSR "Demokritos",
153-10, Aghia Paraskevi, Athens, Greece.}
\date{\today}

\begin{abstract}
In this work we report on the influence of stray fields for both
low and high magnetic fields applied parallel to hybrid trilayers
that consist of a low-T$_c$ Nb interlayer and of two outer
Ni$_{80}$Fe$_{20}$ layers having in-plane but no uniaxial
anisotropy as this is evidenced by complete magnetization data
obtained in the normal state. At low magnetic fields these
trilayered hybrids exhibit a pronounced magnetoresistance effect.
The dynamic transport behavior of the trilayers is presented in
the regime of the magnetoresistance effect through detailed I-V
characteristics. More importantly, the detailed evolution of the
{\it longitudinal} and {\it transverse} magnetic components of the
trilayers is presented from very close to T$_c^{SC}$ to well
inside the superconducting regime. These data clearly show that
below T$_c^{SC}$ and for low magnetic fields the transport
properties of the Nb interlayer are influenced by {\it transverse}
stray-fields that motivate subsequent {\it transverse} magnetic
coupling of the outer Ni$_{80}$Fe$_{20}$ layers. By generalizing
this experimental finding we propose that the generic prerequisite
for the occurrence of intense magnetoresistance effects in
relative ferromagnetic-superconducting-ferromagnetic trilayers is
that the coercive fields of the respective outer layers should
almost coincide. Thus, owing to the {\it simultaneous} occurrence
of magnetic domains all over the surface of the ferromagnetic
layers these are susceptible to {\it transverse} magnetic coupling
mediated by the accompanying {\it transverse} stray fields
occurring mainly above domain walls. Finally, the behavior of the
Ni$_{80}$Fe$_{20}$-Nb-Ni$_{80}$Fe$_{20}$ trilayer's upper-critical
field is investigated in comparison to Nb-Ni$_{80}$Fe$_{20}$
bilayers and Nb single layers. The trilayer's upper-critical field
exhibits a pronounced suppression for low magnetic fields
indicative of a $2D$ behavior, while for high values the
conventional $3D$ behavior represented by a linear temperature
variation is recovered. A similar process is observed in both the
Nb-Ni$_{80}$Fe$_{20}$ bilayer and in the Nb single layer reference
films. However, significant qualitative and quantitative
differences exist between these samples. Based on a mechanism that
is motivated by {\it longitudinal} stray-fields existing
exclusively in the high-field regime we propose a possible
interpretation for this experimental finding.

\end{abstract}


\maketitle

\pagebreak

\section{Introduction}

Since the discovery of the giant magnetoresistance (GMR) effect
\cite{Baibich88} trilayers (TLs) comprised of two ferromagnetic
(FM) electrodes and a normal metal (NM) interlayer have attracted
much interest \cite{Dieny94}. Apart from a NM an insulator (IN)
may also be used as an interlayer so that the hybrid TL is called
tunnel valve and the respective effect tunnel
magnetoresistance.\cite{Parkin98,Parkin00} Recently, relative
hybrids consisting of superconducting (SC) and FM constituents
have attracted much interest both experimentally
\cite{Allsworth02,Gu02,Potenza05,Moraru06,Pena05,Rusanov06,StamopoulosPRB05,StamopoulosPRB06,StamopoulosSST06A,StamopoulosPRB07,Steiner06,Singh07,Visani07,Monton07}
and
theoretically.\cite{Volkov03,Volkov06,Eschrig05L,Eschrig07,Maleki06,Buzdin99,Tagirov99}
The concept of a superconductive spin valve was theoretically
proposed in Refs.\onlinecite{Buzdin99,Tagirov99}. It is based on a
FM-SC-FM TL where the nucleation of superconductivity can be
controlled by the in-plane {\it relative} magnetization
orientation of the outer FM layers. J.Y. Gu et al. \cite{Gu02}
were the first who reported on the experimental realization of a
[Ni$_{82}$Fe$_{18}$-Cu$_{0.47}$Ni$_{0.53}$]/Nb/[Cu$_{0.47}$Ni$_{0.53}$-Ni$_{82}$Fe$_{18}$]
spin valve. I.C. Moraru et al. \cite{Moraru06} also studied a
great number of Ni-Nb-Ni TLs and observed a significantly larger
shift of the superconducting transition temperature T$_c^{SC}$
than that reported in Refs.\onlinecite{Gu02,Potenza05}. In these
works \cite{Gu02,Potenza05,Moraru06} the exchange bias was
employed in order to "pin" the magnetization of the one FM layer
and it was observed that when the magnetizations of the two FM
layers were parallel (antiparallel) the resistive transition of
the SC was placed at lower (higher) temperatures. V. Pe\~{n}a et
al. \cite{Pena05} and A.Yu. Rusanov et al. \cite{Rusanov06},
studied
La$_{0.7}$Ca$_{0.3}$MnO$_3$-YBa$_2$Cu$_3$O$_7$-La$_{0.7}$Ca$_{0.3}$MnO$_3$
and Ni$_{80}$Fe$_{20}$-Nb-Ni$_{80}$Fe$_{20}$ TLs, respectively and
they reported the exact opposite behavior; they observed that when
the magnetizations of the two FM layers were parallel
(antiparallel) the resistive transition of the SC interlayer was
placed at higher (lower) temperatures. In a very recent work of
ours \cite{StamopoulosPRB07} that dealt with exchange biased
Ni$_{80}$Fe$_{20}$-Nb-Ni$_{80}$Fe$_{20}$ TLs we have also derived
the same conclusions. Also, C. Visani et al. in
Ref.\onlinecite{Visani07} confirmed the experimental results of
Ref.\onlinecite{Pena05} only very recently. In addition, in the
work of V. Pe\~{n}a et al. \cite{Pena05} and C. Visani et al.
\cite{Visani07} a GMR-like effect was observed that it was related
to the occurrence of spin imbalance \cite{Takahashi99} in the SC,
ultimately motivated by the high spin polarization (almost
$100\%$) of La$_{0.7}$Ca$_{0.3}$MnO$_3$.

Here we present results on stray-fields motivated features for
both low and high magnetic fields in TLs constructed of low spin
polarized Ni$_{80}$Fe$_{20}$ and low-T$_c$ Nb. These TLs exhibit a
pronounced magnetoresistance (MR) effect ($\Delta
R/R_{nor}\times100\%=45\%$) in the low-field
regime.\cite{StamopoulosSub} Their dynamic transport behavior is
presented through detailed I-V characteristics. These TLs are
plain, in the sense that don't incorporate the mechanism of
exchange bias as the ones studied in
Ref.\onlinecite{StamopoulosPRB07}. Specifically, we present the
complete magnetic characterization of the TLs that they exhibit
in-plane but no uniaxial anisotropy. More importantly, detailed
magnetization measurements of both the {\it longitudinal} and {\it
transverse} magnetic components that are performed from very close
to well below T$_c^{SC}$ reveal the complete evolution of a
magnetostatic coupling of the outer Ni$_{80}$Fe$_{20}$ layers that
forces the Nb interlayer to attain a magnetization component {\it
transverse} to the external magnetic field. These magnetic
characteristics are cornerstones for the explanation of the
observed MR effect since they prove that this is motivated by a
{\it transverse} magnetic interconnection of the outer
Ni$_{80}$Fe$_{20}$ layers through {\it transverse} stray fields
that emerge near coercivity. Ultimately, these {\it transverse}
stray fields locally amount to the lower or even the upper
critical field of the Nb interlayer so that they suppress its
transport properties.

By generalizing our experimental findings we propose a generic
{\it transverse} stray-fields mechanism that could assist us
toward the explanation of the contradictory experimental results
that have been reported in the recent literature regarding the
transport properties of relevant FM-SC-FM
TLs.\cite{Gu02,Potenza05,Moraru06,Pena05,Rusanov06,StamopoulosPRB05,StamopoulosPRB06,StamopoulosSST06A,StamopoulosPRB07,Steiner06,Singh07,Visani07}
According to this proposal the generic prerequisite for the
occurrence of intense MR effects is that the coercive fields of
the respective outer FM layers should almost be the same. When
this condition is fulfilled, the two FM layers are susceptible to
{\it transverse} magnetic coupling that is efficiently mediated by
the {\it transverse} stray fields accompanying the magnetic domain
walls that emerge {\it simultaneously} all over the surface of the
FM layers around their nearly common coercivity.

Finally, the behavior of the TL's upper-critical field is
investigated in comparison to the respective ones of a SC-FM
bilayer (BL) and a SC single layer (SL) reference films. The TL's
upper-critical field, H$_{c2}^{SC}$(T) exhibits an intense
crossover from two-dimensional ($2D$) to three-dimensional ($3D$)
behavior when compared to the BL and the SL. This is exhibited by
a pronounced rounding observed in the H$_{c2}^{SC}$(T) curve for
low magnetic fields, while for high values the conventional linear
temperature variation is recovered. Since for high magnetic fields
the two FM layers are saturated possible {\it transverse} stray
fields located above domain walls are minimized and only {\it
longitudinal} ones that run along their surfaces are mainly
present. Based on a {\it longitudinal} stray-fields mechanism we
propose a possible interpretation for this experimental feature.

\section{Preparation and characterization of samples. Experimental details}

\begin{figure}[tbp] \centering%
\includegraphics[angle=0,width=7.5cm]{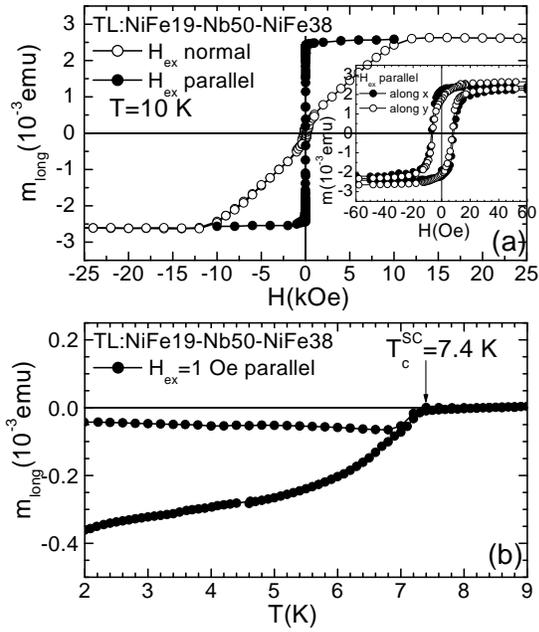}
\caption {(Panel (a)) Longitudinal magnetization m(H) data
obtained for a TL at T$=10$ K$>$T$_c^{SC}$ when the external
magnetic field was both parallel (solid circles) and normal (open
circles). (Panel (b)) Magnetization m(T) curve obtained at
H$_{ex}=5$ Oe applied parallel to the same TL for determination of
its T$_c^{SC}$. The offset owing to the FM layers has been
removed. (Inset) The respective m(H) data shown in the main panel
(a) as obtained when the external field was applied along the two
in-plane dimensions of the TL.}
\label{b1}%
\end{figure}

The samples were sputtered on Si $[001]$ substrates under an Ar
environment ($99.999 \%$ pure). In order to eliminate the residual
oxygen that possibly existed in the chamber we performed Nb
pre-sputtering for very long times since Nb acts as a strong
oxygen
getter.\cite{Stamopoulos05PRB,StamopoulosSST05A,StamopoulosSST06A}
The Nb layers were deposited by dc-sputtering at $46$ W and an Ar
pressure of $3$
mTorr,\cite{Stamopoulos05PRB,StamopoulosSST05A,StamopoulosSST06A}
while for the Ni$_{80}$Fe$_{20}$ (NiFe) layers rf-sputtering was
employed at $30$ W and $4$ mTorr. In this work we show detailed
results for NiFe(19)-Nb(50)-NiFe(38) TLs (in nanometer units).
Quantitatively, the same results were obtained for
NiFe(19)-Nb(50)-NiFe(19) ones. We should stress that: (i) all
depositions were carried out {\it at room temperature} and (ii)
{\it no} external magnetic field was applied during the deposition
of the NiFe layers. However, the samples can't be shielded from
the residual magnetic fields (of the order $10-15$ Oe) existing in
the chamber of our sputtering unit. Thus, our NiFe films exhibit
in-plane anisotropy. However, they don't exhibit uniaxial
anisotropy since the magnetic field sources are placed
symmetrically on the perimeter of the circular rf-gun. The
produced NiFe films have low coercive fields of order $10$ Oe. The
critical temperature of the TLs studied in this work is
T$_c^{SC}\approx 7.4$ K. These characteristics are shown in
figures \ref{b1}(a) and \ref{b1}(b) where the {\it longitudinal}
magnetic component is presented for cases where the external
magnetic field was applied either parallel or normal to the TL.

%------------------------------------------------------------------------
%-------------------------------------------------------------------------

Our MR measurements were performed by applying a dc-transport
current ($I_{{\rm dc}}=0.5$ mA, always normal to the magnetic
field) and measuring the voltage in the standard four-point
straight configuration. The temperature control and the
application of the magnetic fields were achieved in a
superconducting quantum interference device (SQUID) (Quantum
Design). In most cases presented in this work the applied field
was parallel to the films. Cases where it was applied normal will
be specified explicitly.

\section{Transport data}

%------------------------------------------------------------------------
\begin{figure}[tbp] \centering%
\includegraphics[angle=0,width=7.5cm]{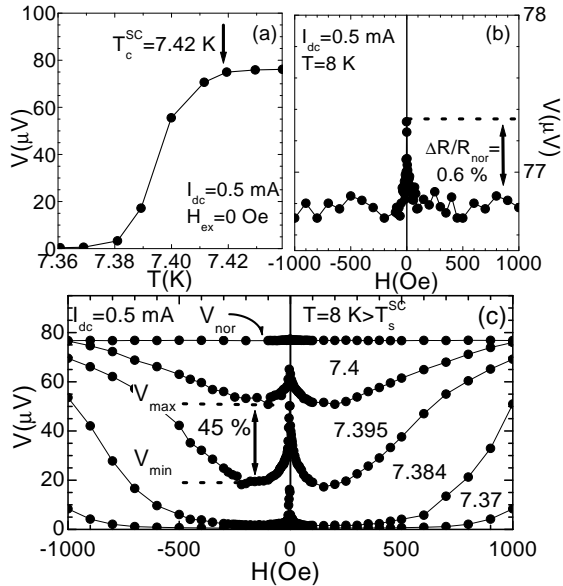}
\caption {(Panel (a)) Zero-field resistive curve V(T) of a TL.
(Panel (b)) V(H) curve at T$=8$ K$>$T$_c^{SC}=7.42$ K. (Panel (c))
V(H) curves at various temperatures across the TL's resistive
transition presented in panel (a). In all cases the magnetic field
was applied parallel to the TL.}
\label{b2}%
\end{figure}%
%-------------------------------------------------------------------------

As may be seen in figure \ref{b2}(a) the zero-field critical
temperature of the TL is T$_c^{SC}=7.42$ K. This value is very
close to the one that is magnetically determined (see figure
\ref{b1}(b)). Figures \ref{b2}(b) and \ref{b2}(c) show the main
raw transport results of our work. Presented are detailed voltage
curves V(H) as a function of magnetic field in the normal (panel
(b)), and in the superconducting (panel (c)) state at various
temperatures across its zero-field resistive curve presented in
panel (a). All these data were obtained for the parallel-field
configuration. We see that for T$>$T$_c^{SC}$ conventional MR is
revealed with values $0.6\%$. In contrast, in the superconducting
state the percentage resistance change
$(R_{max}-R_{min})/R_{nor}\times100\%$ that we observe is
$45\%$.\cite{note} Thus, as we enter the superconducting state the
observed {\it increase is of two orders of magnitude!} This
signifies that the mechanisms motivating the MR peaks in the
normal and superconducting states of the TL hybrid are surely
different. Here we should stress that the pronounced MR effect
discussed for the NiFe-Nb-NiFe TLs is not observed in more simple
NiFe-Nb bilayers (BLs).\cite{StamopoulosSub} In these BLs only a
minor effect exists; the respective percentage resistance change
$(R_{max}-R_{min})/R_{nor}\times100\%$ by no means exceeds $8\%$.
For detailed transport results on the BLs see
Ref.\onlinecite{StamopoulosSub}. According to the quantitative
criteria the MR effect that we observe in the TLs can be
considered as GMR.\cite{Pena05} However, as we show below the main
mechanism motivating the observed phenomenon is not related to the
usual spin-dependent scattering mechanism motivating the GMR
effect.\cite{Baibich88,Dieny94}

%------------------------------------------------------------------------
\begin{figure}[tbp] \centering%
\includegraphics[angle=0,width=7.5cm]{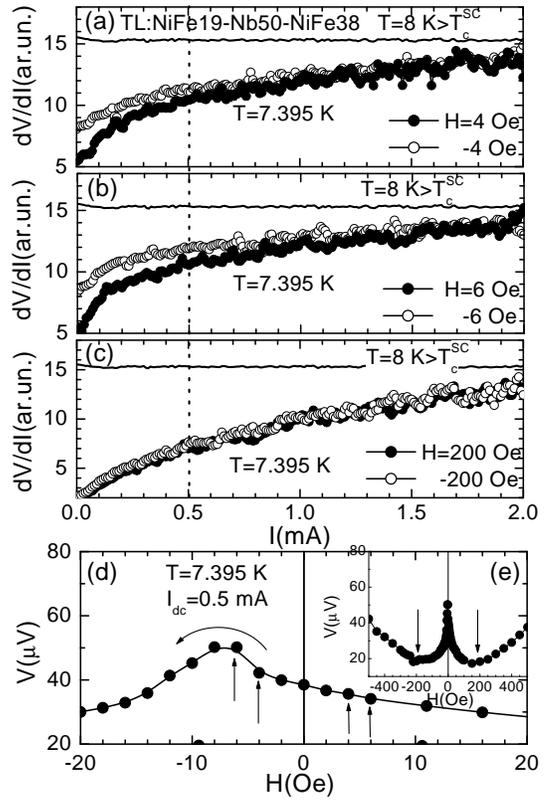}
\caption {Representative I-V characteristics obtained at T$=7.395$
K for various magnetic fields ranging from very close to the MR
peak (Panel (a)-Panel (b)) to far away (Panel (c)). (Panel (d))
Low-field details of the MR V(H) curve obtained at T$=7.395$ K.
The inset presents an extended field range of the MR curve for
clarity. The arrows indicate the exact points where the I-V
experiments were performed. In all cases the magnetic field was
applied parallel to the TL.}
\label{b3}%
\end{figure}%
%-------------------------------------------------------------------------

The dynamic behavior of the MR effect is revealed by the I-V
characteristics shown in figures \ref{b3}(a)-\ref{b3}(e). These
data were obtained at T$=7.395$ K for the parallel-field
configuration. In panels (a)-(b) we present low-field data around
the MR peak, while in panel (c) shown is a representative curve
the reveals the behavior for high magnetic fields away from the MR
peak's maximum. Low-field details of the respective V(H) curve are
presented in panel (d), while its inset shows an extended field
range for clarity. The arrows indicate the exact points where the
I-V characteristics were carried out. The dotted vertical line in
panels (a)-(c) shows the specific value I$_{dc}=0.5$ A of the
transport current where the V(H) curve shown in panel (d) was
obtained. From these data we clearly see that for low magnetic
fields, i.e. around the MR peak the polarity of the applied field
plays a crucial role. In contrast, for high magnetic fields, i.e.
away from the MR peak the obtained I-V curves don't depend on
field's polarity since they clearly coincide. In addition, even
for low magnetic fields the I-V response depends on the transport
current. For low applied currents the curves obtained for opposite
polarity clearly diverge, while for high applied currents they
eventually coincide.

\section{Magnetization data}

First of all we have to underline that our TLs exhibit a very soft
magnetic character of in-plane anisotropy. However, they don't
have uniaxial anisotropy so that as coercivity is approached most
possibly the magnetization reverses in a non-coherent way; the FM
layers segregate into magnetic domains that each one reverses
independently from the others. These characteristics are shown in
figure \ref{b1}(a). In the inset we present representative
magnetization loops that were obtained while the external magnetic
field was applied parallel to the TL and in two different
directions along its two basic dimensions. We clearly see that the
two loops coincide absolutely exhibiting coercivity of less than
10 Oe. The same behavior was also obtained for other directions
were the magnetic field was applied. Thus, we conclude that our
hybrid TLs exhibit in-plane but no uniaxial anisotropy. Finally,
we have to stress that regarding the in-plane anisotropy we can't
distinguish between a component that originates from an intrinsic
mechanism and another one that has a shape origin.

%------------------------------------------------------------------------
\begin{figure}[tbp] \centering%
\includegraphics[angle=0,width=7.4cm]{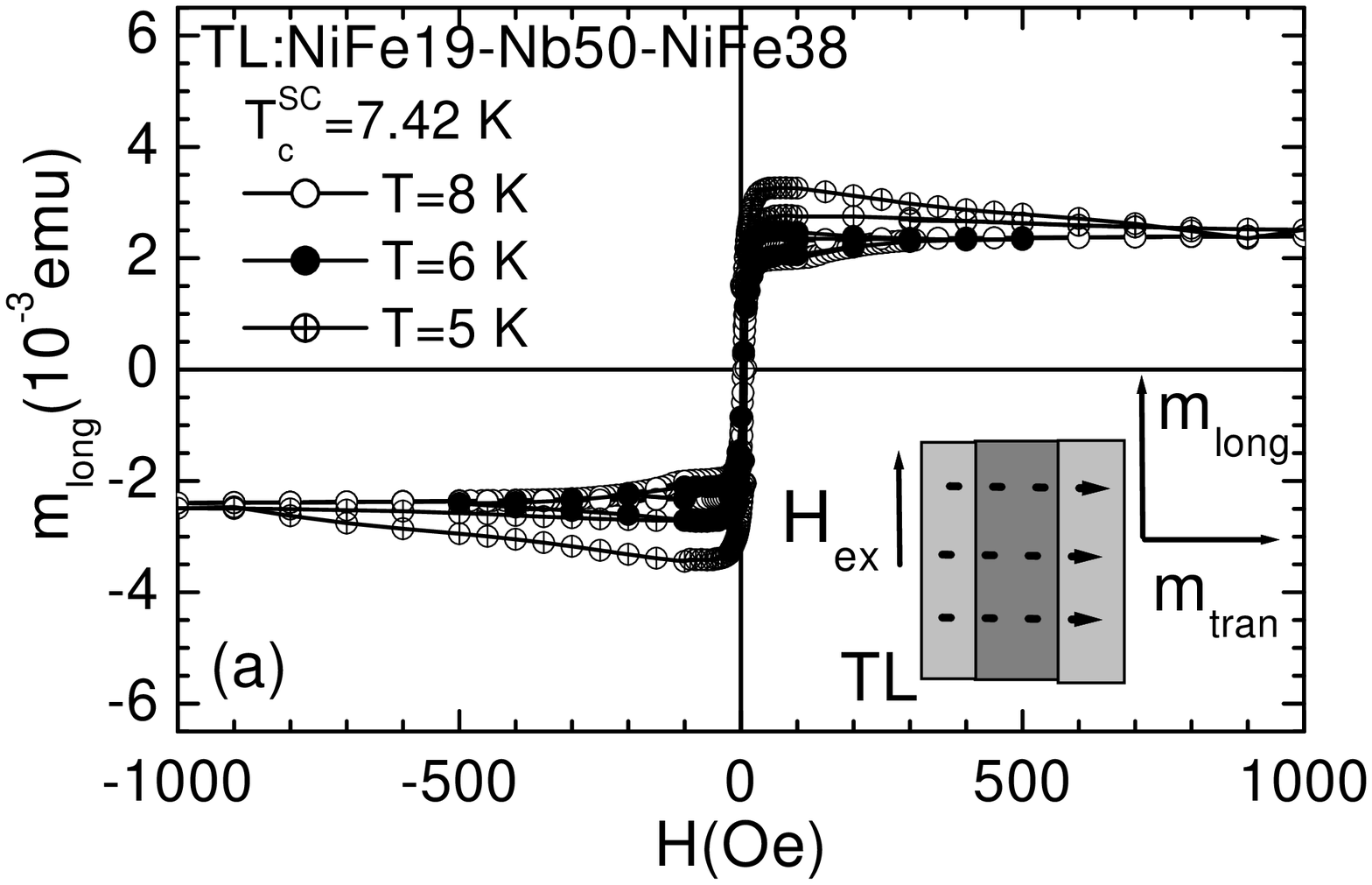}
\includegraphics[angle=0,width=7.5cm]{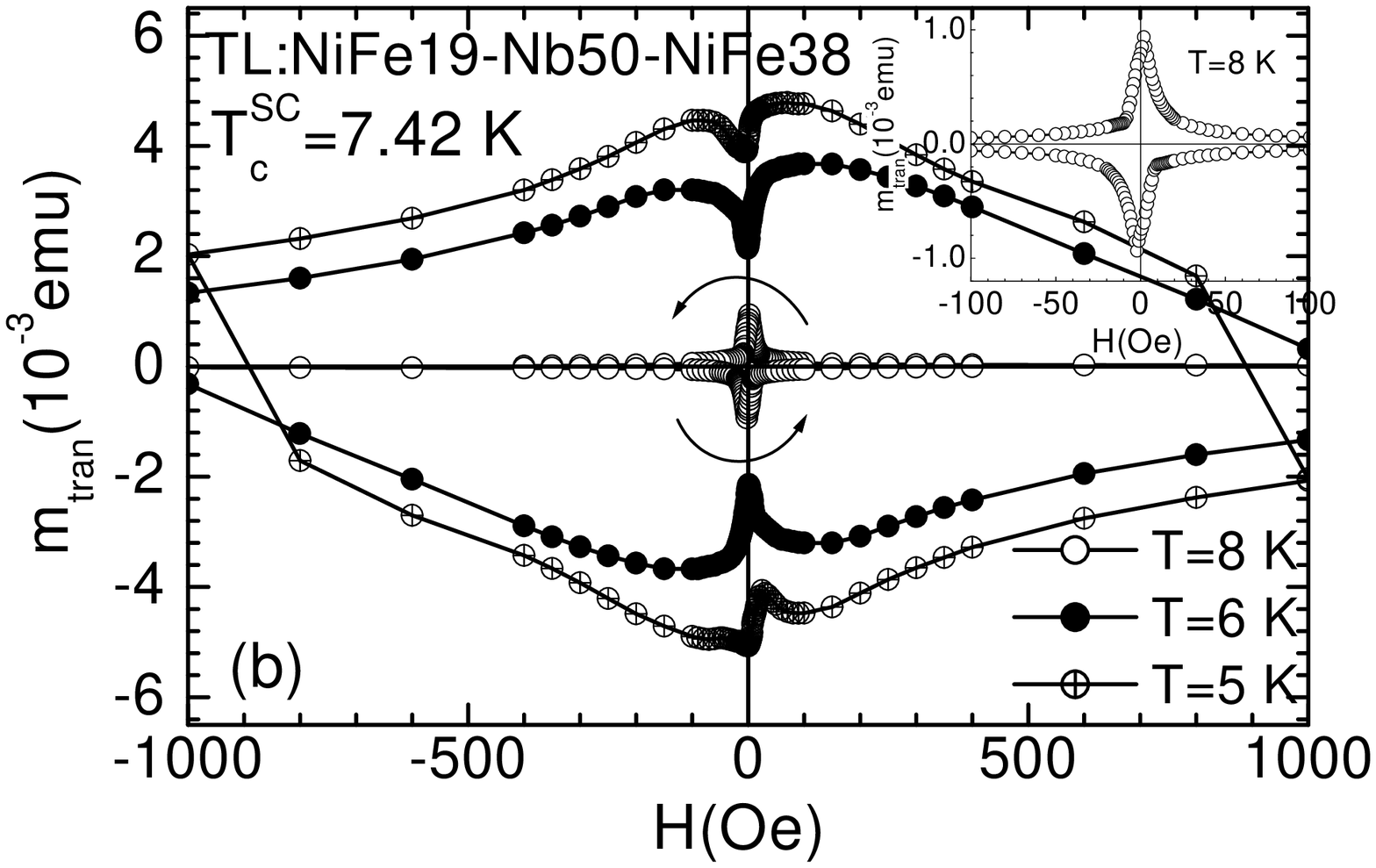}
\caption {(Panel (a)) Magnetization m(H) data referring to the
longitudinal magnetic component of a TL obtained both above and
well below T$_c^{SC}$. (Panel (b)) The respective data for its
transverse magnetic component. The inset of panel (a) presents
schematically the configuration of the external field and of each
magnetic component of the TL, while the respective of panel (b)
focuses on low-field details of the transverse magnetization curve
obtained in the normal state. In all cases the magnetic field was
applied parallel to the TL.}
\label{b4}%
\end{figure}%
%-------------------------------------------------------------------------

In order to investigate the underlying mechanism responsible for
the pronounced MR effect we performed detailed magnetization loops
for the {\it longitudinal} and {\it transverse} magnetic
components both well below and above T$_c^{SC}$. Representative
results are shown in figures \ref{b4}(a) and \ref{b4}(b) for the
parallel-field configuration. The inset of panel (a) presents the
exact configuration of the external field and of each magnetic
component, while the one of panel (b) focuses on low-field details
of the transverse magnetization curve obtained at T$=8$
K$>$T$_c^{SC}$. Let us first discuss panel (a), that is the data
referring to the longitudinal component. These data reveal a
surprising fact: we see that not only above T$_c^{SC}$ but even
well below T$_c^{SC}$ the longitudinal component resembles the
loop of a FM as if the SC is absent (the only fingerprint of its
presence comes from the comparatively small irreversibility that
shows up below T$_c^{SC}$). This reveals that the outer NiFe
layers interact strongly through the Nb interlayer even well
inside the superconducting state so that bulk superconductivity is
strongly suppressed at least when referring to the longitudinal
component. Referring to the transverse component, in panel (b) we
see that the normal-state curve attains significant values of the
order $35\%$ of the saturated longitudinal one (see the respective
inset). This clearly proves that in the normal state a significant
part of the TL's magnetization rotates out of plane near zero
magnetic field. Proceeding with the data obtained in the
superconducting state we stress that they reveal a striking
feature: the out-of-plane rotation of the TL's magnetization is
not observed only in the normal state but also well inside the
superconducting state of Nb. We see that the transverse component
of the TL obtains the model loop expected for a SC when the
mechanism of bulk pinning of vortices dominates.\cite{Tinkham}
{\it The data presented in figure \ref{b4} imply that the SC
behaves diamagnetically not in respect to the parallel external
field but in respect to a new transverse field that emerges owing
to the magnetic coupling of the outer NiFe layers as this is
schematically presented in the inset of panel (a) by the dotted
arrows.}

%------------------------------------------------------------------------
\begin{figure}[tbp] \centering%
\includegraphics[angle=0,width=7.6cm]{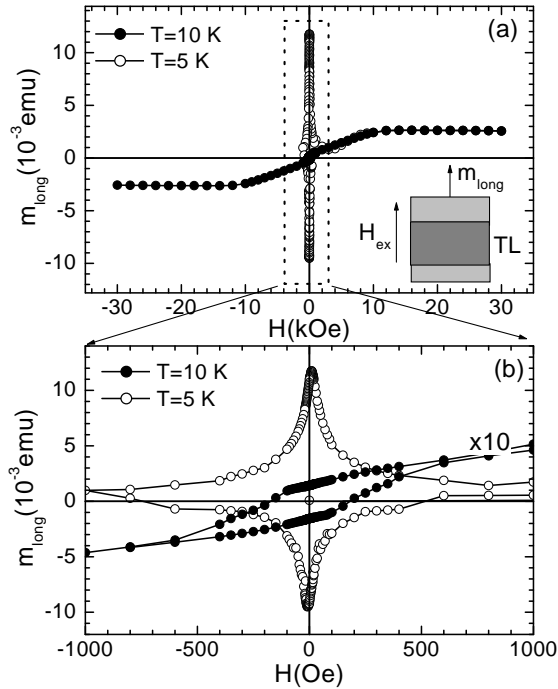}
\caption {(Panel (a)) Longitudinal magnetization m(H) data that
were obtained when the magnetic field was applied normal to the TL
for both above and well below T$_c^{SC}$ in an extended field
range. (Panel (b)) The respective data focused in the low-field
regime. For clarity the curve obtained at T$=10$ K is multiplied
by a factor of $10$. The inset of panel (a) presents schematically
the configuration between the normal magnetic field and the
measured magnetization component.}
\label{b5}%
\end{figure}%
%-------------------------------------------------------------------------

For the normal-field configuration the obtained results are more
conventional. Representative magnetization loops for the {\it
longitudinal} magnetic component for both above and well below
T$_c^{SC}$ for the normal-field configuration are shown in figures
\ref{b5}(a) and \ref{b5}(b) for high and low values of the
magnetic field, respectively. The inset of panel (a) presents
schematically the configuration between the normal magnetic field
and the longitudinal magnetization component. We clearly see that
in this field configuration the magnetization loop that is
obtained below T$_c^{SC}$ resembles the model one expected for a
SC when the mechanism of bulk pinning of vortices
dominates.\cite{Tinkham}

\section{Discussion}

\subsection{Stray-fields {\it transverse} coupling of the outer FM layers}

It is well known that in relevant TLs of non-superconducting
interlayer the interaction of the outer FM layers through stray
fields that occur at domain walls may lead to significant
magnetostatic coupling.\cite{Parkin98,Parkin00} This behavior is
expected to be pronounced when the FM layers have a multidomain
magnetic state, that is in the low magnetic field regime near
coercivity as this is observed in our FM-SC-FM TLs. S. Parkin and
colleagues have shown that such stray-fields coupling that occurs
at domain walls plays a unique role in FM-IN-FM and FM-NM-FM
TLs.\cite{Parkin98,Parkin00} Accordingly, in our case the partial
out-of-plane rotation of the TL's magnetization that we observe in
the normal state (see figure \ref{b4}(a)), and most importantly
{\it its unique transverse magnetic component that we observe in
the superconducting state} (see figure \ref{b4}(b)) are motivated
by such a stray-fields magnetostatic coupling of the outer NiFe
layers. We believe that this is the case since the pronounced MR
effect is observed around the coercive field where the FM layers
segregate in magnetic domains so that a rich reservoir of {\it
transverse} stray fields is available to mediate their magnetic
coupling {\it all over their surface}.

\begin{figure}[tbp] \centering%
\includegraphics[angle=0,width=7.5cm]{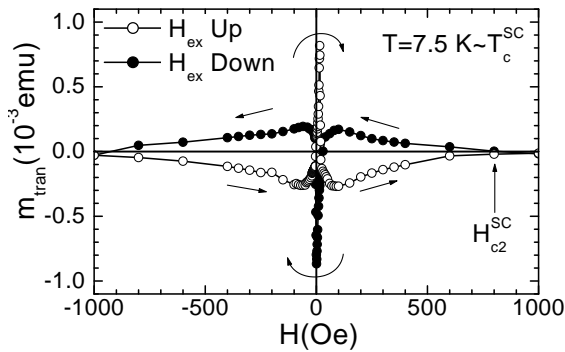}
\caption {Complete magnetization m(H) loop for the transverse
magnetic component of a TL obtained just below its T$_c^{SC}$.}
\label{b6}%
\end{figure}

Ultimately, the magnetostatic coupling discussed here could
motivate the MR peaks presented in figure \ref{b2}(c) as follows:
the {\it transverse} stray fields that interconnect the outer NiFe
layers penetrate completely the Nb interlayer as this is
schematically presented in the inset of figure \ref{b4}(a) by the
dotted arrows. Consequently, in some areas not only the lower but
even the upper critical field of the Nb interlayer may be exceeded
locally. This should result in a suppression of its
superconducting properties for low magnetic fields where this
transverse field gets maximum (see figure \ref{b4}(b)). Indeed,
this is evident not only in our MR curves (see figure
\ref{b2}(c)), but also in our transverse magnetization data where
clear dips are observed near zero field probably indicating a
small suppression of bulk superconductivity (see figure
\ref{b4}(b)). Under a different point of view the data presented
in figure \ref{b4}(b) may be interpreted as following: below
T$_c^{SC}$ the magnetization of the Nb interlayer is {\it
opposite} to the one of the outer NiFe layers, indicating the {\it
diamagnetic} behavior of the SC in respect to the FM layers'
transverse magnetization field. In order to show the clear
evolution of this effect in figure \ref{b6} we present a detailed
loop for another TL very close to the T$_c^{SC}$. We clearly see
that the transverse magnetic component of the SC interlayer is
{\it opposite (diamagnetic)} to the respective of the FM outer
layers.

%------------------------------------------------------------------------
%-------------------------------------------------------------------------

These arguments explains naturally the absence of the pronounced
MR effect in the NiFe-Nb BLs that was stated
above.\cite{StamopoulosSub} Although even in the BLs the
magnetization of the single NiFe layer may partially rotate out of
plane it doesn't have the opportunity to get coupled with an
adjacent NiFe one (as happens in the TLs) so that the existing
{\it transverse} stray fields don't penetrate the Nb interlayer
completely (see below). As a consequence the MR effect observed in
the BLs ($(R_{max}-R_{min})/R_{nor}\times100\%=8\%$) is only minor
when compared to the one of the TLs
($(R_{max}-R_{min})/R_{nor}\times100\%=45\%$).

\subsection{Comparison with relevant experiments}

In this section we compare our results with other experimental
works available in the recent literature. First of all let us
compare our results with those presented in
Ref.\onlinecite{Rusanov06} since both works refer to the same FM
and SC constituents. We believe that the specific magnetic
characteristics that our films attain owing to the preparation
process explain the differences in both the magnitude of the MR
($5 \%-10 \%$ in Ref.\onlinecite{Rusanov06}, while $40 \%-50 \%$
in our work \cite{note}) and the proposed mechanisms between the
TLs studied in Ref.\onlinecite{Rusanov06} and in our work. A. Yu.
Rusanov et al. reported \cite{Rusanov06} on the MR of similar
NiFe-Nb-NiFe TLs and attributed the MR peaks that they observed to
the in-plane antiparallel alignment of the outer NiFe layers.
Their samples exhibited strong in-plane uniaxial anisotropy. Thus,
in their case as the magnetic field passes through zero the
magnetization of each NiFe layer reverses {\it coherently} and
probably entirely in-plane (although data on the transverse
magnetic component are not presented in
Ref.\onlinecite{Rusanov06}). In our case it is surely extreme to
invoke that the magnetization of the outer NiFe layers rotates
coherently; their very soft magnetic character ensures that as
coercivity is approached a rich multidomain magnetic state is
attained. Thus, {\it transverse} stray fields that emerge above
domain walls that are spread all over the surface of the FM layers
may mediate their magnetic coupling efficiently. Unavoidably, the
SC interlayer will be magnetically "pierced".

We stress that the importance of the out-of-plane rotation of the
outer FM layers' magnetizations revealed by our results could be
relevant in all relative FM-SC hybrids that were studied very
recently.\cite{Gu02,Potenza05,Moraru06,Pena05,Rusanov06,StamopoulosPRB05,StamopoulosPRB06,StamopoulosSST06A,StamopoulosPRB07,Steiner06,Singh07,Visani07}
The influence of stray fields originating from a FM multilayer on
an adjacent SC was already highlighted by some of us in
Refs.\onlinecite{StamopoulosPRB05,StamopoulosPRB06,StamopoulosSST06A}.
R. Steiner and P. Ziemann revealed the influence that stray fields
have on a SC interlayer in their recent work \cite{Steiner06}
referring to Co-Nb-Fe and CoO-Nb-Fe TLs for the parallel-field
configuration. Also, very recently C. Monton et al.
\cite{Monton07} studied stray-fields induced effects in Nb/Co
multilayers when the magnetic field was applied parallel to their
surface and reported a behavior similar to the one observed in
Refs.\onlinecite{StamopoulosPRB05,StamopoulosPRB06,StamopoulosSST06A}.
However, C. Monton et al. \cite{Monton07} expanded our knowledge
to the complete mutual effect; the authors \cite{Monton07}
revealed convincingly that the screening coming from the SC layers
influences drastically the magnetic domain state of the adjacent
FM ones. In contrast, A Singh et al. in Ref.\onlinecite{Singh07}
studied in the normal-field configuration Co-Pt/Nb/Co-Pt TLs of
perpendicular magnetic anisotropy and claimed that stray fields
don't actually influence the SC interlayer. This interpretation
was based on experimental data that were obtained when an
insulating SiO$_2$ layer was introduced at each Co-Pt/Nb interface
so that the electronic coupling of the two constituents was
removed. However, in the specific transport experiments that were
presented in Ref.\onlinecite{Singh07} each FM layer was in its
saturated magnetic state so that possible stray fields were only
limited at the specimen's edges.

Here, we present uncontested evidence for the dominant influence
of {\it transverse} stray-fields coupling of the outer NiFe layers
on the Nb interlayer in the respective TLs (see also
Ref.\onlinecite{StamopoulosSub}). We have to stress that in our
work we refer not to the stray fields that exist {\it only near
the edges} of a homogeneously magnetized FM as it was discussed in
Ref.\onlinecite{Singh07} but to the {\it transverse} stray fields
that emerge naturally {\it all over the surface} of the FM owing
to the appearance of magnetic domains around its coercivity. This
is a very important difference that should not be disregarded.

\subsection{Generic prerequisite for the occurrence of pronounced stray-fields {\it
transverse} magnetic coupling }

%------------------------------------------------------------------------
\begin{figure}[tbp] \centering%
\includegraphics[angle=0,width=6.5cm]{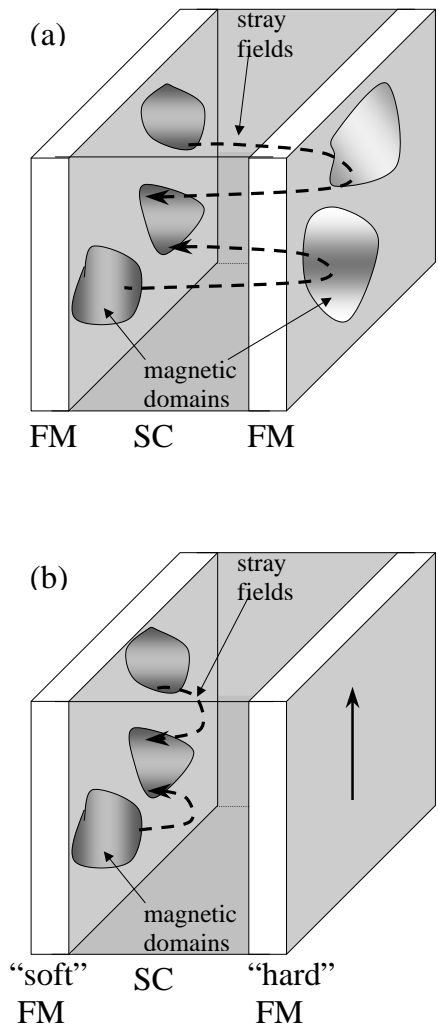}
\caption {Schematic representation of (Panel (a)) efficient and
(Panel (b)) incomplete stray-fields coupling of two FM layers
spaced by a SC interlayer. In the first case the FM layers have
almost identical coercive fields, while in the second one their
coercive fields are quite different.}
\label{b7}%
\end{figure}%
%-------------------------------------------------------------------------

In this subsection we generalize our experimental findings and we
propose a generic {\it transverse} stray-fields mechanism that may
be present in all FM-SC-FM TLs and multilayers. This mechanism is
based on a single prerequisite for the occurrence of {\it
transverse} stray-fields magnetic coupling between FM layers that
are separated by SC interlayers. We propose that in order an
intense MR effect to be observed {\it the coercive fields of the
FM layers should be approximately the same so that the two FM
layers should be susceptible to get magnetically coupled}. This
condition ensures that in the same field range where the first FM
layer exhibits a multidomain magnetic structure that is
accompanied by a rich reservoir of {\it transverse} stray fields
all over its surface, also the second FM layer should be divided
in magnetic domains so that the stray fields outgoing the first
layer will be efficiently sunk in the second one. In the case
where the two FM layers don't share almost common coercive fields
the effectiveness of {\it transverse} stray-fields coupling should
be minimum. For instance, consider the case where the first FM
layer is quite magnetically "soft", while the second one is rather
magnetically "hard". Around its coercivity the "soft" FM layer
will be accompanied by rich stray fields that since they are
randomly distributed will not be effectively delivered to the
"hard" FM layer owing to its robust ordered magnetization. This
scenario is schematically presented in figure \ref{b7}. In panel
(a) we see that when the respective coercive fields coincide the
simultaneous occurrence of magnetic domains in the outer FM layers
will ensure their magnetic coupling so that the SC interlayer
should be magnetically "pierced", completely. In contrast, in
panel (b) we present the case where the respective coercive fields
are very different. Thus, complete coupling of the FM layers can't
be accomplished so that the SC interlayer may repel the weak {\it
transverse} stray fields that originate from each single FM layer
around its coercivity. Consequently, in the first case the
transport properties of the SC interlayer are seriously affected,
while in the second one only a minor influence occurs.

This proposition explains naturally the occurrence of the
pronounced MR peaks that were observed in
Refs.\onlinecite{Pena05,Visani07,Steiner06}. In all these works
the FM layers participating a TL had almost identical coercive
fields as this is evidenced by the only minor two-step behavior
observed in the presented m(H) loops. In contrast, when the
coercive field of the first FM layer is significantly different
than that of the second one the stray-fields {\it transverse}
magnetic coupling is not intense so that the magnetoresistance
effect is getting weak. This is evidenced by the experimental
results presented in
Refs.\onlinecite{Gu02,Potenza05,Moraru06,Rusanov06}. In
Ref.\onlinecite{Rusanov06} the two FM layers exhibit a few tens Oe
different coercive fields as this is clear by the distinct
two-step behavior in the presented m(H) loops. Furthermore, in
Refs.\onlinecite{Gu02,Potenza05,Moraru06} the coercive fields of
the outer FM layers differ by several hundreds Oe. Summarizing
these data, compared to the magnetoresistance effect reported in
Refs.\onlinecite{Pena05,Visani07,Steiner06}, where the FM layers
share almost common coercive fields, the one reported in
Ref.\onlinecite{Rusanov06}, where distinct coercive fields of the
FMs are clearly resolved, is weaker. Eventually, in
Refs.\onlinecite{Gu02,Potenza05,Moraru06}, where the coercive
fields are very different, notable MR peaks were not reported.
Thus, it is natural to assume that when the coercive regimes of
the two FM layers exhibit significant overlapping the observed MR
effect is pronounced, while as the coercivities get progressively
different the MR peaks eventually disappear. {\it We believe that
the only mechanism that could be invoked for the consistent
interpretation of all these experimental data that are reported in
the recent literature
\cite{Gu02,Potenza05,Moraru06,Pena05,Rusanov06,Visani07,Steiner06,StamopoulosSub,StamopoulosPRB07}
is the stray-fields coupling of the outer FM layers that "pierces"
magnetically the SC interlayer.} The specific condition described
here should not be restricted only to TLs of SC interlayer  but
should also hold for TLs having NM and IN
interlayers.\cite{Baibich88,Dieny94,Parkin98,Parkin00}

\subsection{{\it Longitudinal} stray fields and the upper-critical field curve of the TL}

\begin{figure}[tbp] \centering%
\includegraphics[angle=0,width=7.5cm]{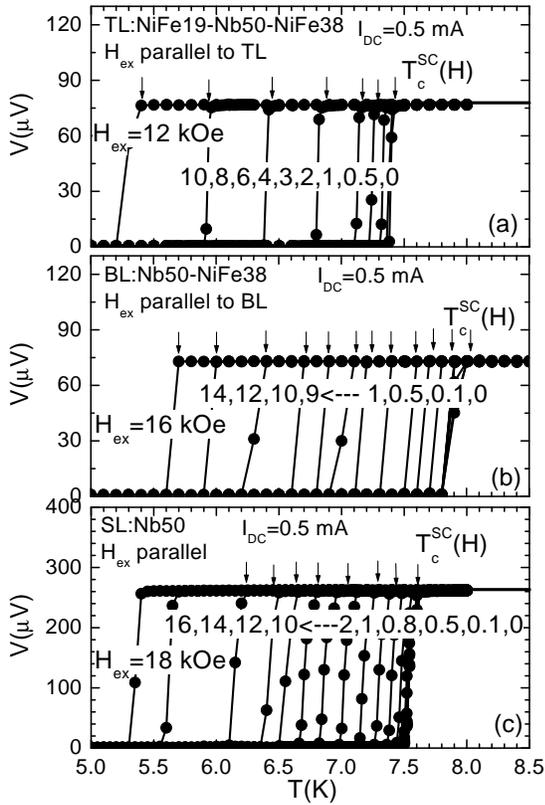}
\caption {Representative raw transport V(T) curves for a
TL:NiFe$19$-Nb$50$-NiFe$38$ (Panel (a)), a BL:Nb$50$-NiFe$38$
(Panel (b)), and a SL:Nb$50$ (Panel (c)). In all cases the
magnetic field was applied parallel to the films and the current
I$_{dc}=0.5$ mA was applied normal to the external field.}
\label{b8}%
\end{figure}

Finally, in this subsection we discuss the intriguing behavior of
the upper-critical field that is observed in the FM-SC-FM TLs for
the parallel-field configuration studied in this work. Also,
comparative results are presented for a SC-FM BL and a SC SL.
Representative raw transport data are shown in figures
\ref{b8}(a)-\ref{b8}(c) for a TL:NiFe$19$-Nb$50$-NiFe$38$, a
BL:Nb$50$-NiFe$38$, and a SL:Nb$50$, respectively.

%------------------------------------------------------------------------
%-------------------------------------------------------------------------

\begin{figure}[tbp] \centering%
\includegraphics[angle=0,width=7.5cm]{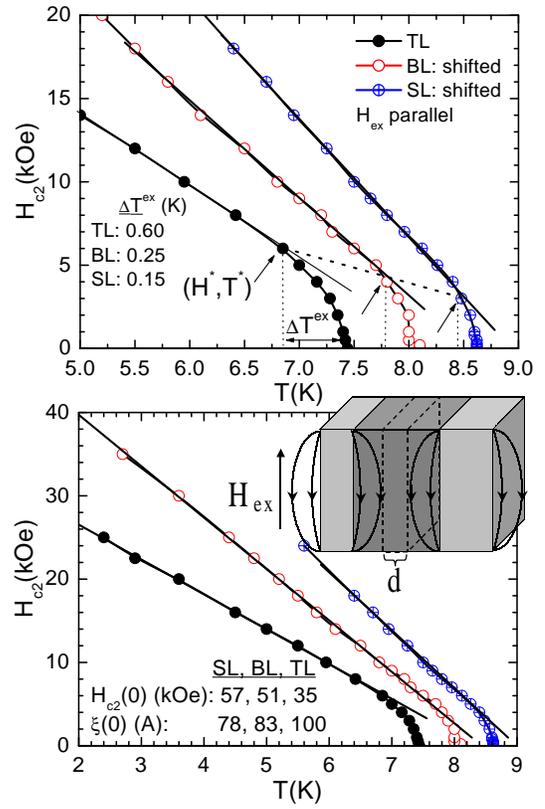}
\caption {Upper-critical field curves for a
TL:NiFe$19$-Nb$50$-NiFe$38$ (solid circles), a BL:Nb$50$-NiFe$38$
(open circles), and a SL:Nb$50$ (circles with crosses) in the
high-temperature, low-field regime (Panel (a)) and in an extended
range (Panel (b)). In all cases the magnetic field was applied
parallel to the films. The BL's and SL's curves are shifted in
temperature for the sake of presentation. The obtained curves
change behavior from $2D$ to $3D$ at different points
(H$^*$,T$^*$) for each specimen. The TL (SL) exhibits the most
(least) wide $2D$ regime. More specifically, at
(H$^*$,T$^*$)$=$($6$ kOe,$6.8$ K) the H$_{c2}^{SC}$(T) curve of
the TL changes behavior from $3D$ (H$>$H$^*$ and T$<$T$^*$) to
$2D$ (H$<$H$^*$ and T$>$T$^*$). The inset presents schematically
the outer FM's {\it longitudinal} stray fields that penetrate the
SC interlayer beside the interfaces. A "stray-fields-free regime"
of thickness $d$ may be formed in the SC's interior (see text for
details).}
\label{b9}%
\end{figure}

In figures \ref{b9}(a)-\ref{b9}(b) we present the respective
upper-critical field curves H$_{c2}^{SC}$(T) for relatively low
and high magnetic fields, respectively. The curves of the BL and
the SL are shifted in temperature for the sake of the
presentation. These data clearly reveal that the hybrid TL
experiences a significant suppression in its upper-critical field
curve in the low-field regime when compared to both the BL and the
reference Nb SL. In contrast, for magnetic fields exceeding $6$
kOe the linear temperature dependence of H$_{c2}^{SC}$(T) is
recovered.

%------------------------------------------------------------------------
%-------------------------------------------------------------------------

The experimental data that are presented in figures
\ref{b9}(a)-\ref{b9}(b) show clearly that for the TL the
H$_{c2}^{SC}$(T) curve experiences an intense crossover from a
$2D$ to $3D$ behavior at (H$^*$,T$^*$)$=$($6$ kOe,$6.8$ K). It
seems that for low magnetic fields and close to T$_c^{SC}$, i.e.
for H$<$H$^*$ and T$>$T$^*$ the Nb interlayer behaves as a $2D$ SC
having thickness lower than the coherence length, $\xi(T)$ of the
Cooper pairs.\cite{Tinkham} In contrast, for H$>$H$^*$ and
T$<$T$^*$ it behaves as a usual $3D$ SC film. The respective BL
and the Nb SL also exhibit such a crossover but of much lower
intensity. In fact the characteristic point (H$^*$,T$^*$) where
the crossover is observed is shifted at progressively lower
(higher) fields (reduced temperatures, T$/$T$_c^{SC}$) starting
from the TL to the SL.

There is no doubt that for the Nb SL the observed crossover
relates exclusively to a thermally driven process that takes place
as the temperature dependent coherence length,
$\xi(T)=\xi(0)[1/(1-T/T_c^{SC})^{1/2}]$ gets equal to the
thickness of the film since it diverges at
T$_c^{SC}$.\cite{Tinkham} Indeed, for the Nb SL the
zero-temperature coherence length $\xi(0)$ as this is estimated
from the zero-temperature extrapolation of the H$_{c2}^{SC}$(T)
curve by using the mean-field theoretical formula $H_{\rm
c2}(T)={\Phi _o/2\pi \xi^2(T)}$ is $\xi(0)\simeq 78$ $\AA$
(H$_{c2}^{SC}(0)\simeq 57$ kOe).\cite{Tinkham} Subsequently, by
simulating the temperature variation of the coherence length
$\xi(T)=\xi(0)[1/(1-T/T_c^{SC})^{1/2}]$ we are able to find the
characteristic temperature T$^*$ where $\xi(T)$ equals the
thickness of the Nb SL. The theoretically estimated temperature
width $\Delta T^{th}=T_c^{SC}-T^*$ where such a $2D$ behavior
should be observed is $\Delta T^{th}\approx 0.2$ K. This value
correlates nicely with the experimental one $\Delta T^{ex}=0.15$
K. In the related Table we present all the experimentally obtained
and the theoretically estimated parameters for all TL, BL and SL.

A number of observations may be extracted from these data. The
experimentally determined temperature widths $\Delta T^{ex}$
exceed significantly the theoretically estimated ones $\Delta
T^{th}$ especially for the TL. In principle both $\Delta T^{ex}$
and $\Delta T^{th}$ should depend on the quality of the SC layer
since the zero-temperature values of both the upper-critical field
H$_{c2}^{SC}(0)$ and the coherence length $\xi(0)$ depend on the
existing crystal disorder (recall that in the dirty limit the
zero-temperature coherence length $\xi(0)$ is limited by the
mean-free path parameter $l$ since the expression $\xi(0)\approx
\sqrt{l\xi_{o}}$ holds, where $\xi_{o}$ is the BCS coherence
length in the clean limit).\cite{Tinkham} Thus, a variation of the
experimentally observed $\Delta T^{ex}$ values among different
samples should not be paradox. However, in our data we observe a
{\it systematic} evolution of the characteristic point
(H$^*$,T$^*$) where the crossover occurs. Moreover, the field
dependence of the crossover point as this is evident by the
gradual increasing of the H$^*$ value when an additional FM layer
is getting added from the SL to the BL and from the BL to the TL
can't be interpreted by the concept of a purely {\it thermally}
driven $2D$ to $3D$ process. Instead, we believe that there is a
{\it field} driven process that contributes, significantly.

%---------------------------------------------------------------------
\begin{table}[tbp] \centering%
\caption{Theoretically estimated and experimentally accessible
values of the temperature width $\Delta T$ where $2D$ behavior
should be observed for samples NiFe$19$-Nb$50$-NiFe$38$ TL,
NiFe$19$-Nb$50$ BL, and NiFe$38$ SL. Also, presented are the
experimental zero-temperature critical field H$_{c2}^{SC}(0)$ and
coherence length $\xi(0)$ values that were used for the
theoretical estimations. Finally, presented are the values of the
characteristic field H$^*$ where the $2D$ to $3D$ crossover is
observed for each sample.}
\begin{ruledtabular}
\begin{tabular}{cccccc}
$  $   &$\Delta T^{th}$ (K) &$\Delta T^{ex}$ (K) &H$_{c2}^{SC}(0)$ (kOe)  &$\xi(0)$ (\AA) &H$^*$ (kOe)\\
$TL$     &0.4    &0.6    &35   &100  &6\\
$BL$     &0.3    &0.25   &51   &83   &4\\
$SL$     &0.2   &0.15   &57   &78    &3\\

\end{tabular}
\end{ruledtabular}
\label{table}
\end{table}
%---------------------------------------------------------------------

Below, based on the influence of the {\it longitudinal} stray
fields of the saturated outer FM layers on the SC interlayer we
discuss such a possible field driven process. Since the saturation
field of the NiFe outer layers is H$_{sat}^{FM}=20$ Oe (see inset
of figure \ref{b1}(a)) even in the low field regime,
H$_{sat}^{FM}<$H$<$H$^*$ where the rounding of the TL's
H$_{c2}^{SC}$(T) curve is observed these FM layers are {\it
saturated}. Thus, since magnetic domains don't exist at saturation
{\it transverse} stray fields are not expected; for
H$_{sat}^{FM}<$H$<$H$^*$ the stray fields are {\it longitudinal}
since they run along the surfaces of the outer NiFe layers (see
the schematic inset of figure \ref{b9}(b)). These {\it
longitudinal} stray fields penetrate the Nb interlayer at a depth
that depends on both the geometric characteristics of the hybrid
structure and the magnetic characteristics of the FM layers (for
instance, the saturation magnetization determines the intensity of
stray fields). However, since the Nb interlayer is quite thick
($50$ nm) we expect that in its interior there could exist a
regime that is not affected by the {\it longitudinal} stray
fields. This is illustrated in the inset of figure \ref{b9}(b) as
a "stray-fields-free regime" of width $d$. Accordingly, for low
magnetic fields the influence of the {\it longitudinal} stray
fields should be considered as the most important parameter that
acts against superconductivity at least at a depth $(50-d)/2$ nm
by the FM-SC interfaces. Especially for magnetically "soft" FM
materials, as is NiFe in our case, these {\it longitudinal} stray
fields could be quite strong ranging between hundreds Oe to a few
kOe. Thus, close to T$_c^{SC}$ where the effect is observed the
H$_{c2}^{SC}$(T) attains low values so that the {\it longitudinal}
stray fields could strongly suppress superconductivity at a depth
of their range. A key-notice is that the {\it longitudinal} stray
fields are opposite to the external magnetic field, H$_{ex}$.
Thus, as the externally applied field increases it should
gradually compensate the {\it longitudinal} stray fields due to
their opposite orientations. Obviously, the compensation will
start from the SC's interior since there the {\it longitudinal}
stray fields are weaker. Consequently, this gradual compensation
will expand the "stray-fields-free regime" and $d$ will increase,
eventually getting equal to the Nb interlayer's thickness as the
external magnetic field equals the intensity of stray fields. {\it
Thus, in order to explain the $2D$ to $3D$ crossover observed in
the TL at (H$^*$,T$^*$)$=$($6$ kOe,$6.8$ K) we speculate that
according to the process described above in low magnetic fields,
H$<$H$^*=6$ kOe the width $d$ of the "stray-fields-free regime" is
lower than the coherence length $\xi$ so that the Nb interlayer
behaves as a $2D$ SC, while at some characteristic value of the
applied magnetic field H$^*=6$ kOe the width $d$ of the
"stray-fields-free regime" will become equal to, and eventually
exceed, the coherence length $\xi$ so that for H$>$H$^*=6$ kOe the
Nb interlayer will start to behave as a $3D$ SC.} {\it For the BL
where a second outer FM layer is missing this process should be
weaker. The "stray-fields-free regime" should be wider for the BL
when compared to the TL so that lower applied fields are needed
for entering the $3D$ regime.} This is in nice agreement to our
experimental findings that are shown in figure \ref{b9} and
summarized in the related Table. Of course, in order this
proposition to be proved definitely, a complete series of TLs
where the thickness of the Nb interlayer is systematically varied
should be studied. This systematic work is in progress. However,
we believe that this draft proposal captures consistently the
experimental data already available.

\section{Conclusions}

In summary, we demonstrated that TLs comprised of low spin
polarized NiFe and low-T$_c$ Nb exhibit special features in both
low and high magnetic fields. For low magnetic fields they behave
as superconducting switches since they exhibit a two orders of
magnitude boost in the MR very close to T$_c^{SC}$. Their dynamic
transport behavior is revealed through I-V characteristics. The
normal state magnetic characterization, but most importantly the
complete evolution of both the {\it longitudinal} and {\it
transverse} magnetic components from very close to well below
T$_c^{SC}$ are presented. These data show that in the
superconducting regime the SC attains a {\it transverse} magnetic
component since it behaves diamagnetically not in respect to the
external magnetic field but in respect to {\it transverse} stray
fields that interconnect the outer NiFe layers. Consequently, the
pronounced MR effect observed in our TLs is due to the suppression
of superconductivity by the {\it transverse} stray fields related
to the domain walls that emerge at coercivity.

A plausible {\it transverse} stray-fields mechanism is proposed
that gives fair explanation for the intense MR effects that are
observed in relevant hybrid TLs. This mechanism is based on a
single, and probably, generic prerequisite: intense MR effect are
observed when the coercive fields of the outer FM layers are
almost identical since under this condition the {\it simultaneous}
occurrence of magnetic domains will provide {\it transverse} stray
fields all over the FMs' surfaces that may mediate effectively
their magnetic coupling.

Finally, the upper-critical field curve H$_{c2}^{SC}$(T) of hybrid
TLs is presented and discussed in comparison to the ones of a BL
and a SL reference films. For the TL a pronounced suppression is
observed in the low-field regime, while the conventional linear
temperature variation is recovered for significantly higher
magnetic fields when compared to the BL and the SL. Based on the
influence of {\it longitudinal} stray fields arising from the
outer NiFe layers when these are saturated we propose a possible
model for the interpretation of this intense $2D$ to $3D$
crossover. According to this explanation the respective process
should be weaker for the BL. This is in agreement to our
experimental results.

{\it Note:} This work is dedicated to Professor S.H. Papadopoulos
who has been a great pedagogue and a sincere humanitarian.

\pagebreak


\begin{references}
\bibitem{note} We note that in agreement to Ref.\onlinecite{Rusanov06}
we prefer to use the definition
$(R_{max}-R_{min})/R_{nor}\times100\%$ (that takes into account
the reference resistance value obtained in the normal state) for
the description of the underlying physics since the alternative
one $(R_{max}-R_{min})/R_{min}\times100\%$ (that takes into
account the minimum value obtained in the superconducting state)
suffers from possible singularities originating from the zeroing
of the measured resistance that naturally occurs in every SC (see
Fig.\ref{b2}(c)).
\end{references}
\end{document}